\documentclass[prl,twocolumn,amssymb]{revtex4}
\usepackage{graphicx}

\begin{document}

\title{Superfluid-Insulator Transition of Strongly Interacting Fermi Gases in Optical Lattices}
\author{Hui Zhai and Tin-Lun Ho}
\affiliation {Physics Department, The Ohio State University, Columbus, Ohio 43210}
\date{\today}
\begin{abstract}
We study a quantum phase transition between fermion superfluid (SF) and band insulator (BI) of fermions in optical lattices.  The destruction of the band insulator is driven by the 
energy gain in promoting fermions from valance band to various conducting bands to form Cooper pairs. 
We show that the transition must take place in rather shallow lattice height, $V_{o}/E_{R}$ between 2.23 and 4.14. The latter is the prediction of mean field theory while the former is the value for opening a band gap. 
As one moves across resonance to the molecule side, the SF-BI transition evolves into the SF-Mott insulator transition of bosonic molecules. 
We shall also present the global phase diagram for SF-Insulator transition for the BCS-BEC family. 
\end{abstract}
\maketitle

There have been strong interests in ultra-cold Fermi gases near Feshbach resonance.  Much of these interests have to do with the non-perturbative nature of the problem and the richness of the phenomena. On one hand, this systems has universal properties at resonance\cite{universal}  difficult to calculate by perturbative means. On the other hand, the ground state turns out to be a robust fermion superfluid connecting the usual BCS state to the Bose-Einstein condensate (BEC) on different sides of the resonance\cite{Jin}. The robustness of this superfluid allows one to study intriguing phenomena such as those related to pseudo-gap and unequal spin population. 
Typically, the strength of a superfluid is reflected in its resistance to perturbation. The upper critical magnetic field of a superconductor is a good example. In the case of neutral fermion superfluids,  this corresponds to the critical rotational frequency at which pairing is  switched off.  Perturbation can also take the form of an optical lattice, in which case, the strength of the superfluid is reflected in the critical lattice height $V_{0}^{\ast}$ which destroys superfluidity.  Such experiment has been performed recently by the MIT group\cite{Ketterle}. 

Since a homogenous Fermi gas can be changed continuously from a BCS to a BEC superfluid as a function of scattering length, the question is how the entire BCS-BEC family is affected by a lattice potential.  Clearly, the answer depends on the  fermion density. 
In the recent MIT experiment\cite{Ketterle}, there are two fermions per site. 
Increasing the lattice height $V_{0}$ on fermion side will lead to a transition from BCS supefluid to a band insulator, whereas on the molecular side, this transition will be from a boson superfluid to a Mott insulator with one boson per site\cite{BHM}.  
While it is the same transition, the properties of the superfluid and insulating phases are changed continuously across the resonance.  At resonance, due to the universal behavior of the system, one expects the critical lattice height $V_{0}^{\ast}$ (measured in recoil energy $E_{R}$) at which superfluid-Insulator (SF-I) takes place to be a universal number, which is a measure of the strength of this robust superfluid.  

It is interesting to note that in most solid state material the band gap ($\sim 1\text{eV}$) is several orders of magnitude larger than the typical pairing energy ($\sim 0.1-1\text{meV}$). The situation is very different in strongly interacting Fermi gases. First of all, the pairing energy is comparable to Fermi energy. Secondly, the band gap is highly tunable. This offers one a unique opportunity to study the competition between pairing and lattice effects. 

In this letter, we shall study the transition between superfluid and insulator near Feshbach resonance by varying lattice height. We shall determine the critical lattice height  $V_{0}^{\ast}/E_{\text{R}}$ and the size of the insulating region in a trap; and discuss the global phase diagram at the end.  Our study is based on mean field theory.  Since the effects of pair fluctuation are weak in the BCS regime but become increasingly strong in the BEC regime, mean field theory is only accurate on the BCS side of the resonance, less accurate near resonance, and becomes  very poor on the BEC side. 
The situation is identical to calculating $T_{c}$ for the crossover family, where fluctuation corrections grow rapidly on the BEC side\cite{Tc}. 
The simplicity of mean field field, however, makes it a useful  
first step to study the phase diagram on the BCS side, provided one realizes its limitations. On the other hand,  we shall see that the mean field theory (in the presence of lattices) provides a surprisingly good result at resonance as it is sufficiently close to a rigorous lower bound.  
Moreover, the qualitative feature provided by the mean field theory is informative enough for one to construct a global phase diagram that extends to the BEC side.  

\begin{figure}[bp]
\begin{center}
\includegraphics[width=9.0cm]
{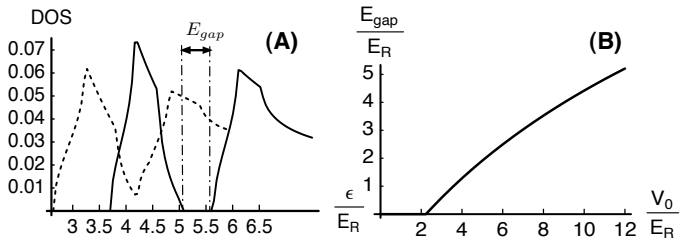}\caption{(A) Density of states (DOS) for $V_0=2E_{R}$ (dashing line, no band gap) and $V_0=3E_{R}$ (solid line, finite band gap) of a cubic lattice.  $E_{\text {gap}}$ is the band gap between the lowest band and the first excited band. (B) Band gap as a function of lattice depth.\label{Band}}
\end{center}
\end{figure}

{\it (A) Band Structure:} The single particle Hamiltonian of a 3D cubic optical lattice is $\mathcal{H}_0=\sum_{i}h_i$,  $i=1,2,3$, $h_i= -\hbar^2\partial^2_{i}/(2M) +V_0\sin^2(Kx_i)$, $K=\pi/\lambda$, and $\lambda$ is the lattice spacing fixed by the wave length of the laser, and $M$ is the mass of the fermion. Typically, $V_{0}$ is measured in unit of 
 ``photon recoil energy" $E_{\text{R}}=\hbar^2 K^2/(2M)$. It is numerically straightforward to calculate 
the Bloch function and energy of the 1D lattice hamiltonian $h$, which we denote as   $\phi_{n, k}$ and $\epsilon_{n,k}$, where $n$ is the band index and $k$ is the crystal momentum. Bloch's theorem implies the expansion $\phi_{n, k}(x) =\sum_m u_{n,k}(m)e^{i(k+mK)x}$, where $k$ lies in the first Brillouin Zone (BZ). The Bloch state of the 3D lattice  $\mathcal{H}_0$ is 
$\phi_{{\bf n}{\bf k}}=\sum_{{\bf G}}u_{{\bf n}{\bf k}}({\bf G})e^{i({\bf k+G}){\bf r}}$, where  ${\bf k} \epsilon BZ$, 
$\phi_{{\bf n}{\bf k}} ({\bf x}) = \prod_{i=1}^{3} \phi_{n_i k_i}(x_{i})$, 
 $u_{{\bf n}{\bf k}}({\bf G})=\prod_{i=1}^{3} u_{n_i k_i}(m_i)$, 
  ${\bf n}=\{n_1,n_2,n_3\}$ is the band index, and ${\bf G}=\{m_1K,m_2K,m_3K\}$ is the reciprocal lattice vector. 
 The energy eigenvalues are  $\epsilon_{{\bf n}{\bf k}}=\sum_{i=1}^{3}\epsilon_{n_i,k_i}$.  

Note that except for the (lowest) $s$-band $(0,0,0)$, all higher bands are degenerate.  For example, the $p$-bands $(1,0,0)$, $(0,1,0)$ and $(0,0,1)$ are three-fold degenerate. Numerically, we found that only  when $V_0/E_{R}>2.23$, a band gap $E_{\text{gap}}$ will develop between the $s$-band and the $p$-bands for this cubic lattice, as shown in 
Fig.\ref{Band}(A).  Fig.\ref{Band}(B) shows how the band gap $E_{\text{gap}}$ grows with lattice height $V_{0}$.

{\it (B) Mean-field Hamiltonian.} The many-body Hamiltonian is $\mathcal{H}=\mathcal{H}_0+\mathcal{H}_{\text{int}}$, $\mathcal{H}_0=\sum_{{\bf n}{\bf k}\sigma}\xi_{{\bf n}{\bf k}}\psi^\dag_{{\bf n}{\bf k}\sigma}\psi_{{\bf n}{\bf k}\sigma}$,  where $\psi^\dag_{{\bf n}{\bf k}\sigma}$ creates fermion with quantum numbers ${\bf n}$, ${\bf k}$ and the spin $\sigma$, $\xi_{\bf n k} = \epsilon_{\bf nk} - \mu$, $\mu$ is chemical potential, and
\begin{equation}
\mathcal{H}_{\text{int}}=g\sum\limits_{{\bf \eta}}\alpha_{{\bf \eta}}\psi^\dag_{{\bf m},{\bf k+p}\uparrow}\psi^\dag_{{\bf m^\prime},{\bf -k+p}\downarrow}\psi_{{\bf n^\prime},{\bf -k^\prime+p}\downarrow}\psi_{{\bf n},{\bf k^\prime+p}\uparrow},
\label{Hint} \end{equation}
where ${\bf \eta}$ denotes the set  $\{ {\bf n,n^\prime,m,m^\prime,k,k^\prime, p}\}$. As the ultraviolet behavior of particle interaction will not be affected by lattice potentias with typical wavelength $\lambda$, the renormlization scheme of the coupling constant $g$ remains to be 
${1}/{g}={m}/{(4\pi\hbar^2 a_{\text s})}-{1}/{\Omega}\sum_{{\bf q}}{m}/{\hbar^2 q^2}$, 
where $a_{{\text s}}$ is the s-wave scattering length, and $\Omega$ is the volume. Here, the ${\bf q}$-sum is over all wavevector ${\bf q}$, not only in the first BZ. 

Pairing with zero crystal momentum is described by terms with  ${\bf p}=0$ in Eq.(\ref{Hint}).
The corresponding coefficient $\alpha_{{\bf \eta}}$ is then 
\begin{eqnarray}
\alpha_{{\bf \eta}}&=&\int d^3{\bf r}\phi^*_{{\bf m},{\bf k}}({\bf r})\phi^*_{{\bf m^\prime},{\bf -k}}({\bf r})\phi_{{\bf n^\prime},{\bf -k^\prime}}({\bf r})\phi_{{\bf n},{\bf k^\prime}}({\bf r})\nonumber\\
&=&\sum\limits_{{\bf G}}Q^{*{\bf G}}_{{\bf m,m^\prime,k^\prime}}Q^{{\bf G}}_{{\bf n,n^\prime,k}}
\end{eqnarray}
where 
$Q^{{\bf G}}_{{\bf n,n^\prime,k}}=\sum_{{\bf Q}}u_{{\bf n},{\bf k}}({\bf G-Q})u_{{\bf n^\prime},{\bf -k}}({\bf G+Q})$, and ${\bf Q}$ is a reciprocal lattice vector. The reduced BCS Hamiltonian then assumes the form
\begin{eqnarray}
&&{\cal H}_{\text{m-f}}=\sum\limits_{{\bf n}\sigma, {\bf k}\epsilon BZ}(\epsilon_{{\bf n,k}}-\mu)\psi^\dag_{{\bf n}{\bf k}\sigma}\psi_{{\bf n,k}\sigma}+\nonumber\\ 
&&\sum\limits_{{\bf G}}(\Delta^*_{{\bf G}}\sum\limits_{{\bf n,n^\prime}, {\bf k}\epsilon BZ}Q_{{\bf n,n^\prime,k}}\psi_{{\bf n,k}\uparrow}\psi_{{\bf n^\prime,-k}\downarrow}+\text{h.c.}-\frac{|\Delta_{{\bf G}}|^2}{g}),\nonumber
\end{eqnarray}
where $
\Delta^*_{{\bf G}}=g\sum_{{\bf n,n^\prime},{\bf k}\epsilon BZ}Q^{*{\bf G}}_{{\bf n,n^\prime,k}}\langle\psi^\dag_{{\bf n,k}\uparrow}\psi^\dag_{{\bf n^\prime,-k}\downarrow}\rangle. 
$
By construction,  ${\cal H}_{\text{m-f}}$ reduces to the non-interacting Hamiltonian ${\cal H}_{0}$ as all $\Delta_{\bf G}$ vanish. The insulator described by this mean field theory is therefore taken by assumption as a band insulator. This is the limitation of the theory and will have to be corrected in a more accurate theory. 

Within this mean field theory, the superfluid to insulator (SF-I) transition is a consequence of the competition between the band insulator and a BCS superfluid. Starting from a band insulator, it is necessary for the pairing interaction to be strong enough to overcome the band gap,  so that a pair of fermions from the lowest band can be promoted to a higher band to facilitate number fluctuation necessary for pairing in different bands. 
 Alternatively, as the band gap increases, it will be more advantageous for a superfluid with density appropriate to a set of filled bands to eventually collapse into these filled bands to minimize its energy through the band gaps.

{\it (C) SF to Insulator transition.} Since we are interested in the onset of pairing order, we can expand the free-energy in powers of $\Delta_{{\bf G}}$, which gives
\begin{equation}
\mathcal{F}=\sum_{{\bf G}}(-\frac{m}{4\pi\hbar^2 a_{\text{s}}}-W_{{\bf G}})|\Delta_{{\bf G}}|^2+o(|\Delta|^4),
\end{equation}
where  $W_{{\bf G}}$ is the pairing susceptibility. At  zero-temperature, it is 
\begin{eqnarray}
W_{{\bf G}}= & \Omega^{-1} \sum\limits_{{\bf k}\epsilon BZ} \sum  \limits_{{\bf n,n^\prime}}\frac{|Q^G_{{\bf n,n^\prime,k}}|^2[\Theta(\xi_{{\bf n,k}})+\Theta(\xi_{{\bf n^\prime,-k}})]}{\xi_{{\bf n,k}}+\xi_{{\bf n^\prime, -k}}}\nonumber\\
  & - \Omega^{-1} \sum\limits_{{\bf q}} \frac{1}{\hbar^2 q^2/m} ,
\end{eqnarray}
where $\Theta(x)=1/2(-1/2)$ for $x>0(x<0)$.
The system is SF if at least one of $\Delta_{{\bf G}}$ is not zero, which requires at least one  $W_{{\bf G}}> 
-m/(4\pi\hbar^2 a_{\text s})$. 

\begin{figure}[bp]
\begin{center}
\includegraphics[width=8.5cm]
{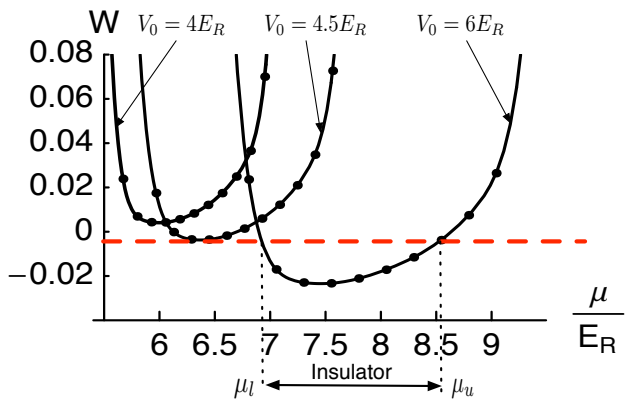}\caption{The function ${\cal W}$ defined in Eq.(\ref{def-W}) as a function of chemical potential $\mu/E_{{\text R}}$, for $V_0=4E_{\text{R}}$, $4.5E_{{\text R}}$ and $6E_{\text{R}}$, respectively. If ${\cal W}$ is above the horizontal red dashing line whose value is  
$-1/(8\pi Ka_{\text{s}})$, the system is a superfluid for all chemical potentials.  Insulator occurs when a portion of ${\cal W}$ falls below $-1/(8\pi Ka_{\text{s}})$. In the case of $V_{o}=6E_{R}$, this occurs in the chemical potential interval $(\mu_{l}, \mu_{u})$.\label{susceptibility}}
\end{center}
\end{figure}

Since  $u_{{\bf n,-k}}(-{\bf Q})$ $=$ $u^*_{{\bf n,k}}({\bf Q})$, we write 
$Q^{G}_{{\bf n,n^\prime,k}}$ $=$ $\sum_{{\bf Q}}$ $u^*_{{\bf n^\prime,k}}({\bf -Q-G})$ $u_{{\bf n,k}}({\bf G-Q})$$=\int d^3{\bf r}$ $\phi^*_{{\bf n^\prime,k}}({\bf r})$ $\phi_{{\bf n,k}}({\bf r})$ $e^{-i2{\bf G}{\bf r}}$. We also find numerically  that $W_{{\bf G=0}}$ is always larger than $W_{{\bf G\neq 0}}$.  So the Cooper instability first occurs at the zero-momentum channel.  Since  $Q^{\bf G=0}_{{\bf n,n^\prime, k}}=\delta_{{\bf n,n^\prime}}$, 
the condition for superfluidity can be greatly simplified as 
\begin{equation}
W_{\bf G=0}  = \frac{1}{\Omega}\left(\sum\limits_{{\bf k}\epsilon BZ}\sum\limits_{{\bf n}}\frac{1}{2|\xi_{{\bf n,k}}|}-\sum_{\bf q} \frac{1}{\hbar^2 k^2/m} \right)> - \frac{m}{4\pi\hbar^2 a_{\text s}} .
\label{W}\end{equation}
Eq.(\ref{W}) shows that $W_{\bf G=0}$ must be of the form 
\begin{equation}
W_{\bf G=0} = K^3 E_{R}^{-1} {\cal W}(\mu/E_{R}, V_{0}/E_{R}),\label{def-W}
\end{equation}
where ${\cal W}$ is a dimensionless function. The condition Eq.(\ref{W}) for superfluidity then becomes
\begin{equation}
{\cal W}(\mu/E_{R}, V_{0}/E_{R}) > - 1/(8\pi Ka_{s}). 
\label{calW} \end{equation}

The function ${\cal W}$ has the following properties:  

\noindent (a) It diverges whenever $\mu$ lies in a band, and is finite when $\mu$ is within a band gap.  This is a consequence of the integrand $|\xi_{\bf n,k}|^{-1} = 
|\epsilon_{\bf n,k}-\mu|^{-1}$ in Eq.(\ref{W}).  The behavior of ${\cal W}$ for different $V_{o}$ is plotted as a function $\mu$ in Fig.\ref{susceptibility}, where we have only displayed the part of  ${\cal W}$  for $\mu$ within the first band gap\cite{higher}. 

For given $V_{0}$, as $\mu$ rises from the top of the $s$-band to the bottom of the $p$-band, 
${\cal W}$ drops from a divergent value to a finite minimum and then rises back to a divergent value. 
This minimum will occur at $\mu^{\ast}(V_{0})$ somewhere inside the band gap. This minimum value of ${\cal W}$ will be denoted as $ {\cal W}^{\ast}(V_{0})= {\cal W}(\mu^{\ast}(V_{0})/E_{R},  V_{0}/E_{R})$, which is important for determining the phase diagram.  

\noindent (b) As $V_{o}$ increases, the band gap is widened. At the same time, as shown in 
 Fig.\ref{susceptibility},  the curve ${\cal W}$ as well as its minimum $\mu^{\ast}(V_{0})$ are shifted to higher of $V_{0}$, while the value of ${\cal W}^{\ast}(V_{0})$ drops.  
For given $a_{s}$, there is an ``onset"  lattice height (denoted as $V_{\text{onset}}$) where  the minimum of  ${\cal W}$ barely touches $-1/(8 \pi Ka_{s})$,  i.e. 
 $ {\cal W}^{\ast}(V_{\text{onset}}) = {\cal W}(\mu^{\ast}(V_{\text{onset}}) /E_{R},  V_{0}/E_{R}) = -1/(8\pi Ka_{s})$. The chemical potential where this touching occurs will be  denote as   $\mu_{\text{onset}}\equiv \mu^{\ast}(V_{\text{onset}})$. The word ``onset" means that this is the minimal value of $V_{0}$ for band insulator to occur in some range of chemical potential.  

\begin{figure}[tbp]
\begin{center}
\includegraphics[width=8.8cm]
{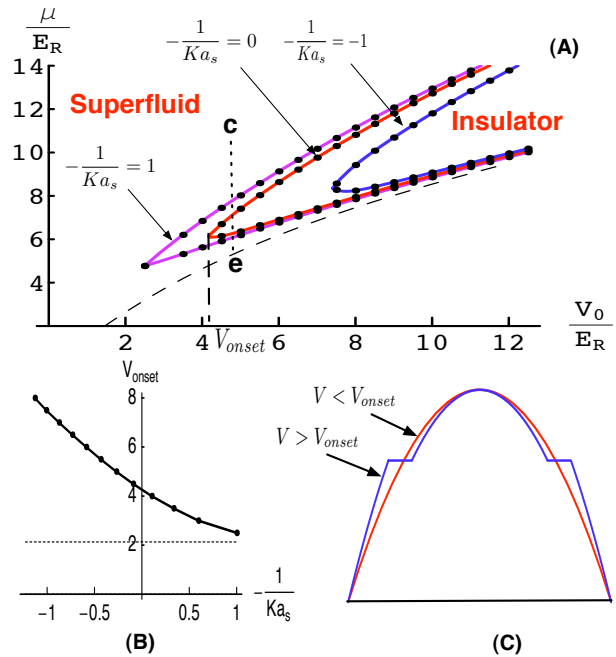}\caption{ (A) The $\mu-V_0$ phase diagram for different scattering length, where $-1/(Ka_{\text{s}})=1$, $0$ and $-1$ respectively. The dashing line denotes the bottom of the lowest band, which monotonically increases as the increase of zero-point energy. The dotted vertical line ${\cal T}$ is the trajectory of local chemical potential form the center of the trap ({\bf c}) to the edge of the cloud ({\bf e}). (B) Solid line: $V_{\text{onset}}$ as a function of $-1/(Ka_{\text{s}})$. Dashing line: the lowest lattice depth for opening up a band gap. (C) Schematic density profile in trap for $V>V_{\text{onset}}$ and $V<V_{\text {onset}}$ respectively. \label{Phasediagram}}
\end{center}
\end{figure}

{\em (D) Determination of phase boundary:} The phase boundary on the $\mu-V_{0}$ plane 
separating the superfluid and the band insulator is given by ${\cal W}(\mu/E_{R}, V_{0}/E_{R}) = - 1/(8\pi Ka_{s})$, which can be determined graphically as shown in Fig.2.   As discussed before, we shall focus on $1/(Ka_{s})< 1$, i.e. on the BCS side and around resonance but not on the BEC side.  The graphical method works as follows: 

\noindent (i)  If ${\cal W}$ intersects $-1/(8 \pi Ka_{s})$ so that it lies below  $-1/(8 \pi Ka_{s})$
over an interval of $\mu$, say, $(\mu_{\ell}, \mu_{u})$, 
as shown in the case of $V_{0}=6E_{\text{R}}$ in Fig.2, the system is a band insulator within this interval of $\mu$. 
See also Fig.\ref{Phasediagram}. 

\noindent (ii) As $V_{0}$ decreases, the curve ${\cal W}$ rises. As a result, the interval $(\mu_{\ell}, \mu_{u})$ shrinks, and the values  $\mu_{\ell}$ and $\mu_{u}$ get closer to each other. When $V_{0}$ reaches a critical value $V_{\text{onset}}$, the insulating interval shrinks to zero, and $\mu_{\ell}=\mu_{u}= \mu_{\text{onset}}$, as shown in the case of $V_0=4.5E_{\text{R}}$. In this way, the points $\mu_{\ell}$ and $\mu_{u}$ trace out a phase boundary shown in Fig.\ref{Phasediagram}. These values $V_{\text{onset}}$ and $\mu_{\text{onset}}$ are identical to those defined in point (b) in Section $(C)$.  In Fig.3, the phase boundaries for different $1/(Ka_{s})$ have been displaced.  The onset lattice height $V_{\text{onset}}$ as a function of $a_{s}$ has also shown in Fig.\ref{Phasediagram}(B). 
At resonance, it shows that $V_{\text{onset}}= 4.14E_{R}$. Since a band gap only appears for $V_{0}> 2.23E_{R}$, the mean field theory therefore predicts that {\em in a three dimensional cubic lattice, the critical lattice height for the onset of the first insulating phase is between $2.23E_{R}$ and $4.14E_{R}$ at resonance}. It should be noted that these numbers predicted here depend on the lattice type (such as f.c.c. and b.c.c.), since they have different band structure. It should also be noted that the lattice height $4.14E_{R}$ for SF-I transition is rather shallow. 
At this lattice height, tight binding approximation is not valid and also the system can not be described by a single band attractive Hubbard model.

\noindent (iii) In the presence of a trap, within the local density approximation, the chemical potential is a function of position, 
$\mu({\bf r}) = \mu-V({\bf r})$, where $V({\bf r})$ is the trapping potential. Going from the center to the surface of the gas, a vertically downward trajectory ${\cal T}$ is generated in the $\mu$-$V_0$ space. If this trajectory passes through the insulating region as shown in Fig.\ref{Phasediagram}(A), this will show up as a step structure in the {\it in-situ} density profile as shown in 
Fig.\ref{Phasediagram}(C). 

\begin{figure}[tbp]
\begin{center}
\includegraphics[width=8cm]
{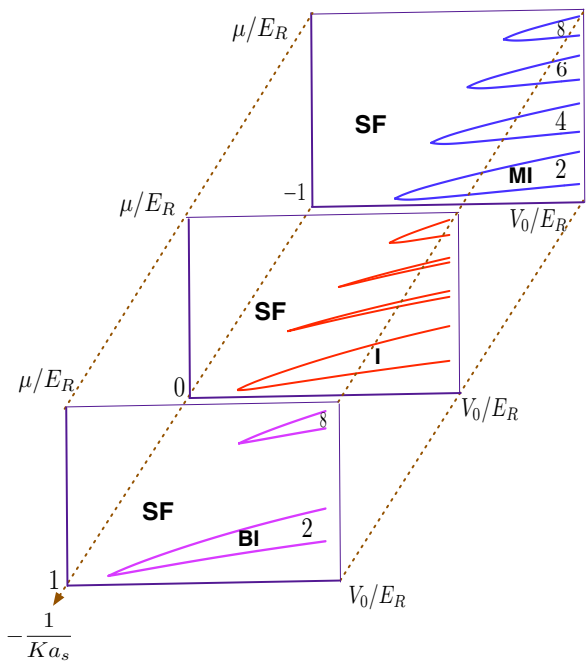}\caption{Schematic of how the $\mu-V_0$ phase diagram evolves across a Feshbach resonance, from the molecule side (up one) to resonance regime (middle one) to the fermion side (bottom one).\label{globalphasediagram}}
\end{center}
\end{figure}
Starting from a superfluid, as $V_{0}$ increases, an insulating lobe will move toward the physical trajectory ${\cal T}$ from the right in Fig.\ref{Phasediagram}(A).   When the lobe touches ${\cal T}$, which happens when $V_{0} = V_{\text{onset}}$ with the point of contact occurs at $\mu_{\text{onset}}$, the insulating phase will develop at the location ${\bf r}$ where $\mu({\bf r}) = \mu_{\text{onset}}$.

{\it (E) Global phase diagram:}  It is useful to consider the phase diagam in the space of chemical potential $\mu$, lattice height $V_{0}$, and interaction parameter $-1/(Ka_{s})$. 
In the absence of lattice, the ground state can evolve continuously from a BCS state to a BEC by shrinking the size of the pair wave function continuously\cite{Leggett}. In the case of two fermion per site, as $V_{o}$ increases, the fermion superfluid on the BCS side (which are made up of Cooper pairs of Bloch states) will evolve continuously into a bosonic superfluid with one boson per site on the average on the BEC side.  For sufficiently high 
$V_{o}$, the ground state on the BCS side  will be a band insulator, whereas it will be a Mott insulator (or a Fock state) with one boson per site on the BEC side. Since a band insulator can be written as a Fock state with two fermions in the same Wannier state on each site. A band insulator can also evolve continuously into a Mott insulator by simple changing the wavefunction of a fermion pair with opposite spin on the same site. In this way, there is a crossover of both superfluid and insulator as one crosses the resonance.   Note that while filling the lowest ($s$) band on the BCS side requires two fermions per site, filling an additional ($p$) band requires totally eight fermions per site.  
The Mott insulators on the BEC side, however, can have any integer number of bosons, or any even integer number of fermions per site.  
To go from the BEC side to the BCS side, only Mott phases with 1 and 4 bosons can crossover to a band insulator. Mott phases with 2 and 3 bosons (or 4 and 6 fermions) per site must disappear as it approaches the fermion side. The phase diagram will therefore look like Fig.\ref{globalphasediagram} schematically. 

This work is supported by NSF Grant DMR-0426149 and PHY-05555576.


\begin{thebibliography}{99}
\bibitem{higher}
For the lattice depths displayed in Fig.(\ref{susceptibility}), either the higher band gaps have not been opened, or if they are opened, the corresponding minima of ${\cal W}$ in these band gaps are higher than that in the lowest band gap, and therefore we do not display them.
\end{thebibliography}
\end{document}